\documentclass[aps,prd,reprint,nofootinbib,superscriptaddress,floatfix]{revtex4-2}
\usepackage{amsmath,amssymb,bm}
\usepackage{graphicx}
\usepackage{xcolor}
\usepackage[normalem]{ulem}
\usepackage[colorlinks=true,allcolors=blue]{hyperref}
\makeatletter
\providecommand\@dotsep{5}
\makeatother                      
\newcommand{\dd}{\mathrm{d}}
\newcommand{\mpl}{M_{\rm Pl}}
\newcommand{\rhoc}{\rho_{\rm c}}
\newcommand{\Lc}{\mathcal{L}_{\rm cusc}}

\begin{document}
\onecolumngrid
\setcounter{tocdepth}{1}
%\hbox{}              
\clearpage
\twocolumngrid
\setcounter{tocdepth}{0}
\title{A Cuscuton Representation of the Loop Quantum Cosmology Bounce}
\author{Niayesh Afshordi}
\email{nafshordi@pitp.ca}
\affiliation{Department of Physics and Astronomy, University of Waterloo,
Waterloo, Ontario N2L 3G1, Canada}
\affiliation{Waterloo Centre for Astrophysics, University of Waterloo,
Waterloo, Ontario N2L 3G1, Canada}
\affiliation{Perimeter Institute for Theoretical Physics, 31 Caroline Street N,
Waterloo, Ontario N2L 2Y5, Canada}

\author{Kristina Giesel}
\email{kristina.giesel@fau.de}
\affiliation{Institute for Quantum Gravity, Theoretical Physics III,
Friedrich-Alexander-Universit\"at Erlangen-N\"urnberg,
Staudtstra{\ss}e 7, 91058 Erlangen, Germany}
\date{\today}

\begin{abstract}
Loop Quantum Cosmology (LQC) replaces the big bang singularity of the
homogeneous universe by a bounce, usually described by the modified
Friedmann equation \(H^2=\rho(1-\rho/\rhoc)/(3\mpl^2)\).  We show that
this background dynamics follows from a local cuscuton effective theory,
whose scalar equation is a constraint rather than a wave equation. One way to establish this is to write the cuscuton in terms of an angular coordinate \(\theta\), identified
with the LQC polymerization angle \(2\lambda b\).  Its constraint gives
\(H\propto\sin\theta\), while the Einstein constraint gives
\(\rho=\rhoc\sin^2(\theta/2)\), exactly reproducing the LQC bounce.  To our
knowledge, this is the first closed-form, local, generally covariant
realization of  the exact standard flat-FLRW LQC background
dynamics for minimally coupled matter satisfying null energy condition,
without introducing additional local dynamical degrees of freedom.
A branchwise Legendre transformation establishes a canonical
equivalence between the clock-gauge-fixed homogeneous mimetic system in the case of vanishing mimetic dust energy density and the cuscuton
systems. It selects the constant-tension cuscuton coordinate
\(\Theta=\sin\theta-\theta\), while \(\theta\) retains its interpretation
as the LQC polymerization angle. This equivalence does not establish
agreement of the inhomogeneous theories or their perturbations. The angular
action also admits an explicit branched \(f(K)\) representation, where
\(K\) is the mean extrinsic curvature of spatial hypersurfaces. The
construction is therefore an effective covariant representation of the
LQC holonomy correction, not a derivation from full loop quantum gravity.
\end{abstract}
\maketitle

\section{Introduction}
Loop quantum cosmology (LQC) gives a controlled symmetry-reduced arena in
which the big bang singularity of the spatially flat FLRW model is
replaced by a bounce \cite{Ashtekar:2006wn,Ashtekar:2011ni}.  At the
level of effective homogeneous dynamics, the key modification is simple:
the connection appears through holonomies, and the Friedmann equation is
replaced by
\begin{equation}
  H^2=\frac{\rho}{3\mpl^2}
  \left(1-\frac{\rho}{\rhoc}\right),
  \label{eq:lqc_friedmann_intro}
\end{equation}
where \(\rhoc\) is the critical density.  The correction is usually read
as a quantum-geometric modification of the gravitational Hamiltonian, not
as exotic matter: ordinary matter can still obey the usual continuity
equation while \(H\) vanishes at \(\rho=\rhoc\).

This paper asks a narrower effective-field-theory question.  Can the
homogeneous LQC constraint be written as the constraint of a local theory
with no additional propagating scalar?  The key observation is that LQC
already supplies 
a natural angular parametrization: the polymerization angle.  We
construct an angular cuscuton whose constraint traces the LQC ellipse in
the \((H,\rho)\) plane and identify \(\theta=2\lambda b\), with \(b\) the
isotropic connection variable.
The coordinate \(\theta\) makes the holonomy origin manifest, while
the constant-tension coordinate \(\Theta=\sin\theta-\theta\) makes the
canonical cuscuton structure manifest.
The cuscuton is suited to this role
because its square-root kinetic term is nondynamical: its field equation
fixes the mean curvature of a preferred foliation rather than propagating
a wave mode \cite{Afshordi:2006ad}.  This also connects the construction
to constant-mean-curvature (CMC) formulations and the low-energy
foliation language of Ho\v rava--Lifshitz gravity
\cite{Afshordi:2009tt,Afshordi:2010de}.

Covariance without a new local mode is highly restrictive.  In four
dimensions, Lovelock's theorem excludes a nontrivial local, metric-only
modification with second-order field equations \cite{Lovelock:1971yv}.
Palatini \(f(R)\) gravity evades this conclusion through an auxiliary
connection, equivalent to a constrained Brans--Dicke scalar at
\(\omega=-3/2\) \cite{Olmo:2008nf,Olmo:2011uz}. In particular,  \cite{Olmo:2008nf} used this structure to reconstruct the massless-scalar LQC background, although the
exact Palatini function is obtained implicitly from a numerical inverse
problem and its closed analytic form is an interpolation.  More recently,
\cite{Delhom:2023loop} constructed a family of metric-affine actions with
accurate analytic fits to LQC and modified-LQC backgrounds with asymmetric bounce. Order-reduced
higher-curvature actions provide another covariant description, but
remove the extra modes of the unreduced theory only perturbatively
\cite{Sotiriou:2008rp}.  
The
angular cuscuton  provides a closed-form, local,
generally covariant action that reproduces
Eq.~\eqref{eq:lqc_friedmann_intro} for minimally coupled matter
with \(\rho+p>0\) on its regular branch,
while adding no local propagating fields.

Cuscuton limiting-curvature cosmologies were developed in \cite{Quintin:2019limiting}. Under the assumptions of
single-field reconstruction within viable Horndeski gravity, \cite{Miranda:2024bounce}
also identifies cuscuton and extended-cuscuton theories as covariant
carriers of effective bounces. Our contribution
is the explicit angular LQC action, its holonomy identification, and its
branchwise Legendre and \(f(K)\) representations.

Limiting-curvature mimetic gravity provides an important neighboring
construction: a multivalued function of the d'Alembertian of a constrained
clock field reproduces the same flat-FLRW LQC ellipse
\cite{Chamseddine:2016uef,Langlois:2017hlu,deHaro:2018sqw,deCesare:2018drf}.
We show below that its nonlinear sector admits a branchwise first-order
form which reduces precisely to the angular cuscuton on a homogeneous
clock slicing.  This gives a direct origin for the otherwise rather
special factors in the cuscuton action.

This relation ends at the homogeneous background.  Standard mimetic
gravity carries a dustlike scalar mode with \(c_s^2=0\), while the
cuscuton carries no additional local scalar and is formally characterized
by \(c_s^2\to\infty\)
\cite{Afshordi:2006ad,Arroja:2015wpa}.  Their agreement on the background
cannot therefore imply perturbative equivalence.  After deriving the
angular action, we expose its relation to mimetic gravity, give its
branched \(f(K)\) form, and assess what would be required to extend the
construction beyond the background.

\section{LQC effective dynamics with holonomy corrections}

We first recall the effective LQC structure that we want to reproduce.
In the improved-dynamics prescription, the isotropic connection variable
\(b\) appears through the polymerization  angle \(\lambda b\)  in the effective Hamiltonian.
Here \(\lambda\) is the polymerization length, with
\(\lambda^2=\Delta\) the minimal area. In units
\(\mpl^2=1\), the effective Hamiltonian constraint is
\begin{equation}
{\cal H}=v(-\rhoc\sin^2(\lambda b)+\rho)\approx 0,
\end{equation}

Here \(v\) is the volume variable canonically paired with \(b\),
with \(\{b,v\}=\gamma/2\) in these units.  The Hamiltonian constraint
yields
\begin{equation}
  \rho=\rhoc\sin^2(\lambda b),
  \qquad
  \rhoc=\frac{3}{\gamma^2\lambda^2},
  \label{eq:lqc_density_holonomy}
\end{equation}
where \(\gamma\) is the Barbero--Immirzi parameter. Hamilton's equation for the
 volume yields
\begin{equation}
  H=\frac{1}{2\gamma\lambda}\sin(2\lambda b).
  \label{eq:lqc_hubble_holonomy}
\end{equation}
Combining \eqref{eq:lqc_density_holonomy} and
\eqref{eq:lqc_hubble_holonomy} gives
\begin{equation}
  3H^2=\rho\left(1-\frac{\rho}{\rhoc}\right),
  \label{eq:lqc_friedmann_units}
\end{equation}
or equivalently \eqref{eq:lqc_friedmann_intro} after restoring \(\mpl\).

Equations \eqref{eq:lqc_density_holonomy}--\eqref{eq:lqc_hubble_holonomy}
show that the LQC bounce is naturally parametrized by a compact
gravitational angle.  Define
\begin{equation}
  \theta\equiv 2\lambda b .
  \label{eq:theta_lqc_def}
\end{equation}
Then
\begin{equation}
  \rho=\rhoc\sin^2\frac{\theta}{2},
  \qquad
  K\equiv 3H=\frac{\sqrt{3\rhoc}}{2}\sin\theta .
  \label{eq:lqc_angle}
\end{equation}
This parametrization is the clue for the effective reconstruction below:
we seek a nondynamical field whose constraint enforces precisely this
polymerization ellipse related to the holonomy corrections.

\section{Angular Cuscuton Reconstruction}

Consider Einstein gravity, ordinary matter, and a cuscuton angle.  In
the main derivation we set \(\mpl^2=1\), use the metric signature
\((-+++)\), and define
\begin{equation}
  X_\theta
  \equiv\sqrt{-g^{\mu\nu}\partial_\mu\theta\partial_\nu\theta}.
\end{equation}
The cleanest form of the model is
\begin{equation}
  \Lc
  =-\sqrt{\frac{\rhoc}{3}}\,(1-\cos\theta)X_\theta
    +\frac{\rhoc}{4}(1-\cos\theta)^2.
  \label{eq:theta_cuscuton}
\end{equation}
This is the usual cuscuton square-root kinetic term, but with a
field-dependent tension.  The field is compact, and the full homogeneous
LQC cycle is naturally covered by \(0\le\theta\le 2\pi\).  The branch
\(\dot\theta<0\) runs from a contracting low-density phase through
\(\theta=\pi\), where the bounce occurs, to the expanding low-density
phase.

The action \eqref{eq:theta_cuscuton} has no quadratic kinetic term.  Its
homogeneous equation of motion is a first-order constraint.  Writing
\begin{equation}
  A(\theta)=\sqrt{\frac{\rhoc}{3}}\,(1-\cos\theta),
  \qquad
  W(\theta)=\frac{\rhoc}{4}(1-\cos\theta)^2,
\end{equation}
the Lagrangian is \(\Lc=-A X_\theta+W\).  On the branch
\(\dot\theta<0\), the \(\theta\) equation gives
\begin{equation}
  3H A(\theta)=W_{,\theta}
  =\frac{\rhoc}{2}(1-\cos\theta)\sin\theta .
  \label{eq:theta_constraint}
\end{equation}
Therefore
\begin{equation}
  H=\frac{1}{2}\sqrt{\frac{\rhoc}{3}}\,\sin\theta,
  \qquad
  K\equiv 3H=\frac{\sqrt{3\rhoc}}{2}\sin\theta .
  \label{eq:theta_hubble}
\end{equation}
This is already the LQC polymerization relation \eqref{eq:lqc_angle}.  The
derivation applies for \(0<\theta<2\pi\).  At \(\theta=2\pi n\), both
\(A\) and \(W_{,\theta}\) vanish and the cuscuton equation becomes
degenerate; the LQC branch reaches its zero-density endpoints by
continuity, but the action also admits separate degenerate GR sectors with
\(\theta\) fixed at those values.

The cuscuton energy density is independent of \(\dot\theta\):
\begin{equation}
  \rho_\theta
  =\dot\theta\frac{\partial\Lc}{\partial\dot\theta}-\Lc
  =-W(\theta)
  =-\frac{\rhoc}{4}(1-\cos\theta)^2 .
  \label{eq:theta_density}
\end{equation}
The Einstein constraint is then
\begin{equation}
  3H^2=\rho-\frac{\rhoc}{4}(1-\cos\theta)^2,
  \label{eq:friedmann_theta}
\end{equation}
where \(\rho\) is the ordinary matter density.  Combining
\eqref{eq:theta_hubble} and \eqref{eq:friedmann_theta} fixes
\begin{equation}
  \rho
  =\frac{\rhoc}{4}\left[\sin^2\theta+(1-\cos\theta)^2\right]
  =\rhoc\sin^2\frac{\theta}{2}.
  \label{eq:rho_theta}
\end{equation}
Equations \eqref{eq:theta_hubble} and \eqref{eq:rho_theta} immediately
give
\begin{equation}
  3H^2=\rho\left(1-\frac{\rho}{\rhoc}\right),
  \label{eq:lqc_reconstructed}
\end{equation}
which is exactly the effective LQC Friedmann equation.

The matter continuity equation
\(\dot\rho=-3H(\rho+p)\), with p the matter pressure, together with the cuscuton constraint, also fixes
the orientation and rate along this ellipse:
\[
  \dot\theta=-\sqrt{\frac{3}{\rhoc}}(\rho+p)
             =-\gamma\lambda(\rho+p)=2\lambda\dot b.
\]
The relation extends through the bounce by continuity. More generally,
either orientation obeys \(|\dot\theta|=\sqrt{3/\rhoc}(\rho+p)\).
Thus a regular solution requires \(\rho+p>0\). Therefore, this construction does not reproduce null-energy-condition
(NEC)-violating LQC
trajectories.

For comparison with standard cuscuton notation, two alternative field
coordinates are useful.  A nonangular density coordinate is
\begin{equation}
  \phi\equiv\rhoc\sin^2\frac{\theta}{2}.
  \label{eq:theta_redefinition}
\end{equation}
On a monotonic half-cycle this rewrites \eqref{eq:theta_cuscuton} as
\begin{equation}
  \Lc=-\mu^2(\phi)
  \sqrt{-g^{\mu\nu}\partial_\mu\phi\partial_\nu\phi}
  -V(\phi),
  \label{eq:general_cuscuton}
\end{equation}
with
\begin{align}
  V(\phi)&=-\frac{\phi^2}{\rhoc}, \label{eq:potential_phi}\\
  \mu^2(\phi)
  &=\frac{2}{\sqrt{3}\,\rhoc}
    \left(\frac{1}{\phi}-\frac{1}{\rhoc}\right)^{-1/2}.
  \label{eq:mu_phi}
\end{align}
Thus the \(\phi\)-form is algebraically convenient, but the \(\theta\)
form makes the polymerization origin of the construction manifest.

\begin{figure}[tbp]
  \centering
  \includegraphics[width=0.78\linewidth]{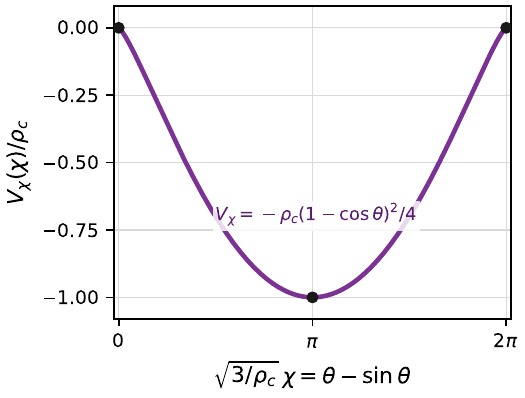}
  \caption{The cuscuton potential term in the canonical field variable,
  plotted in units \(\rhoc=1\).  The horizontal axis is
  \(\sqrt{3/\rhoc}\,\chi=\theta-\sin\theta\), and the plotted function
  is \(V_\chi[\chi(\theta)]/\rhoc=-(1-\cos\theta)^2/4\), the negative potential appearing in Eq.~\eqref{eq:vchi_potential}.}
  \label{fig:vchi_potential}
\end{figure}

The canonical cuscuton normalization is obtained from
\begin{equation}
  \frac{\dd\chi}{\dd\theta}
  =\sqrt{\frac{\rhoc}{3}}\,(1-\cos\theta),
\end{equation}
which integrates to
\begin{equation}
  \chi=\sqrt{\frac{\rhoc}{3}}\,(\theta-\sin\theta).
  \label{eq:chi_theta}
\end{equation}
The action becomes
\begin{align}
  \Lc&=-\sqrt{-g^{\mu\nu}\partial_\mu\chi\partial_\nu\chi}
       -V_\chi(\chi),
       \label{eq:canonical_chi_action}\\
  V_\chi[\chi(\theta)]
  &=-\frac{\rhoc}{4}\left[1-\cos\theta\right]^2 .
       \label{eq:vchi_potential}
\end{align}
Here \(\theta(\chi)\) is defined implicitly by \eqref{eq:chi_theta}.
The \(\theta\)-variable is therefore the clean explicit field, while
\(\chi\) gives the canonical cuscuton normalization.  The negative
potential \eqref{eq:vchi_potential} is shown in
Fig.~\ref{fig:vchi_potential}.

\section{Relation to Limiting-Curvature Mimetic Gravity}
\label{sec:mimetic_relation}

Limiting-curvature mimetic gravity provides a useful cross-check of the
reconstruction.  A branchwise Legendre transformation both derives the
seemingly special angular factors in Eq.~\eqref{eq:theta_cuscuton} and
makes precise why the correspondence stops at the homogeneous background.
The same flat-FLRW dynamics was reconstructed in
Refs.~\cite{Chamseddine:2016uef,Langlois:2017hlu,deHaro:2018sqw}.  In our
\((-+++)\) convention, we introduce a constrained mimetic clock \(T\)
through
\begin{align}
  S_{\rm mim}&=\int\dd^4x\sqrt{-g}\,\mathcal L_{\rm mim},
  \notag\\
  \mathcal L_{\rm mim}
  &=\frac{R}{2}+\mathcal L_m+\frac{\rhoc}{2}F(Y)
  \notag\\
  &\quad+\ell\left(g^{\mu\nu}\partial_\mu T
  \partial_\nu T+1\right),
  \label{eq:mimetic_action}
\end{align}
where $\ell$ is a Lagrange multiplier and
\begin{equation}
  Y\equiv\frac{\Box T}{\Xi_m},
  \qquad \Xi_m\equiv\frac{\sqrt{3\rhoc}}{2}.
  \label{eq:mimetic_y}
\end{equation}
The temporal gauge fixing, \(T=-t\), is convenient for the LQC trajectory used
above: the mimetic constraint fixes the lapse $N$ to unity, and
\(\Box T=3H=K\).  Hence \(Y=\sin\theta\).

The LQC-reconstructing function is multivalued.  It is most cleanly
defined by using \(\theta\) itself as the branch label,
\begin{equation}
  F[\sin\theta]
  =1+\frac{1}{2}\sin^2\theta
   -\theta\sin\theta-\cos\theta .
  \label{eq:mimetic_F_branch}
\end{equation}
Thus neither the inverse sine nor
\(\sqrt{1-\sin^2\theta}\) in the usual expression for \(F(Y)\) is to be
read as a single principal branch.  The branches and their matching
conditions are part of the definition of the limiting-curvature theory
\cite{deCesare:2018drf}.

Introduce the branchwise Legendre variable and its potential,
\begin{align}
  P&\equiv F_{,Y}=\sin\theta-\theta,
  \label{eq:mimetic_legendre_P}\\
  \mathcal V_{\rm L}(P)&\equiv PY-F(Y)
  =-\frac{1}{2}(1-\cos\theta)^2 .
  \label{eq:mimetic_legendre_potential}
\end{align}
The nonlinear sector then has the first-order representation
\begin{align}
  \frac{\rhoc}{2}F(Y)
  &\longleftrightarrow
  \sqrt{\frac{\rhoc}{3}}\,P\Box T
  -\frac{\rhoc}{2}\mathcal V_{\rm L}(P)
  \notag\\
  &=\sqrt{\frac{\rhoc}{3}}\,P\Box T
  +\frac{\rhoc}{4}(1-\cos\theta)^2 .
  \label{eq:mimetic_first_order}
\end{align}
Varying  with respect to \(P\) gives \(Y=\dd\mathcal V_{\rm L}/\dd P\) and recovers the
original branch of \(F\).  Up to a boundary term, the derivative term is
\begin{equation}
  \sqrt{\frac{\rhoc}{3}}P\Box T
  \doteq
  \sqrt{\frac{\rhoc}{3}}(1-\cos\theta)
  \nabla_\mu\theta\nabla^\mu T .
  \label{eq:mimetic_mixed_term}
\end{equation}
For homogeneous fields, \(T=-t\), and the branch \(\dot\theta<0\),
\begin{equation}
  \nabla_\mu\theta\nabla^\mu T
  =\dot\theta=-|\dot\theta|=-X_\theta .
  \label{eq:mimetic_clock_reduction}
\end{equation}
Equations \eqref{eq:mimetic_first_order}--
\eqref{eq:mimetic_clock_reduction} therefore reduce exactly to the
angular cuscuton action \eqref{eq:theta_cuscuton}.  Constant shifts of
\(P\) between inverse-sine branches multiply \(\Box T\) and change the
action only by a boundary term.

This derivation also states the limit of the correspondence.  Before the
homogeneous clock reduction, \eqref{eq:mimetic_mixed_term} is a mixed
two-field derivative operator, not the cuscuton norm.  The mimetic
constraint fixes the norm of \(\nabla_\mu T\), but does not force
\(\nabla_\mu\theta\) to remain parallel to it in an inhomogeneous
spacetime.  The Legendre transformation therefore proves a branchwise
background relation, not a covariant equivalence of the mimetic and
cuscuton theories.  This distinction is consistent with the failure of a
general polymer-LQC interpretation in anisotropic models
\cite{Bodendorfer:2018csn} and with broader comparisons of mimetic and
cuscuton limiting-curvature theories \cite{Sakakihara:2020rdy}.
In particular, \cite{Sakakihara:2020rdy} formulate both theories in
terms of limited extrinsic curvature, identify the extra mimetic
matter-like integration constant, and extend the cuscuton-type theory to
limit anisotropy. Their construction is related
to, but distinct from, the angular LQC reconstruction here.

The homogeneous correspondence can also be made canonical.  Let
\(\alpha\equiv\sqrt{\rhoc/3}\) and
\(U(P)\equiv-\rhoc(1-\cos\theta)^2/4\).  With the clock and orientation
conventions used above, the gauge-fixed homogeneous first-order action
is, up to a boundary term and common matter terms,
\begin{equation}
  S_{\rm first}^{\rm gf}
  =\int\dd t\left[-3a\dot a^2-\alpha a^3\dot P-a^3U(P)\right].
  \label{eq:canonical_first_order}
\end{equation}
The point transformation
\begin{equation}
  \Theta=P=\sin\theta-\theta,
  \qquad P_\Theta=\pi_P,
  \label{eq:canonical_map}
\end{equation}

preserves the canonical one-form,
\(P_\Theta\dd\Theta=\pi_P\dd P\).  In the conventions of
Eq.~\eqref{eq:canonical_first_order}, that is $\dot{\theta}<0$ and $T=-t$, it maps the primary constraints into one another
\begin{align}
  C_{\rm first}^{(1)}&=\pi_P+\alpha a^3\approx0,
  \notag\\
 C_{\rm cusc}^{(1)}&=P_\Theta+\alpha a^3
  =C_{\rm first}^{(1)}\approx0.
  \label{eq:canonical_primary_map}
\end{align}

Writing the cuscuton potential in the $\Theta$ coordinate as $V_\Theta(\Theta) = V_\chi(-\alpha\Theta) = U(\Theta)$, preservation of the primary constraint maps the secondary constraints into one another.  The total Hamiltonians
also coincide with the same Lagrange multiplier.  Thus the two  gauge-fixed \((N=1)\) homogeneous systems have the same reduced constraint structure and Hamiltonian evolution.

Gauge fixing \(N=1\) at the action level removes the equation obtained by varying the lapse, i.e. the Hamiltonian constraint, from the reduced variational problem. The common reduced Hamiltonian is therefore conserved, but its value is not fixed by the gauge-fixed equations alone. To identify the reduced solutions with solutions of the original theory, one must separately impose the condition inherited from the ungauge-fixed system. Including the common matter Hamiltonian \(a^3\rho\), the reduced Hamiltonian is
\[
H_{\rm red}=-\frac{\pi_a^2}{12a}+a^3[U(P)+\rho],
\qquad
\pi_a=-6a\dot a.
\]
On the cuscuton side, the condition inherited from the ungauge-fixed theory imposes \(H_{\rm red}=0\). On the mimetic side, the conserved clock charge instead allows
\[
H_{\rm red}=-C,
\]
where \(C\) is an integration constant, so that
\[
3H^2=\rho+U(P)+\frac{C}{a^3}.
\]
For the same ordinary matter density \(\rho\), the physical homogeneous correspondence therefore requires \(C=0\), corresponding to the limit of vanishing mimetic dust energy density.

The variable \(\Theta\) is a constant-tension cuscuton
coordinate with the opposite orientation to \(\chi\) in
Eq.~\eqref{eq:chi_theta}:
\begin{equation}
  \chi=-\alpha\Theta=-\alpha P.
  \label{eq:chi_legendre_map}
\end{equation}

Accordingly, \(\theta=2\lambda b\) is the natural polymerization-angle
coordinate, while \(\Theta=\sin\theta-\theta\) is the Legendre-dual
cuscuton coordinate in terms of which we can recover the standard cuscuton action. The canonical map $(\Theta=P, P_\Theta=\pi_P)$ is
globally invertible, but its auxiliary parametrization by \(\theta\) is
only branchwise regular because
\(\dd\Theta/\dd\theta=\cos\theta-1\) vanishes at
\(\theta=2\pi n\).  This establishes canonical equivalence between the
clock-gauge-fixed homogeneous mimetic and cuscuton systems in the limit of vanishing dust energy in the mimetic sector. It does not
establish canonical equivalence with LQC or between the full
inhomogeneous theories.

\section{f(K) Representation and Branch Structure}

\begin{figure}[tbp]
  \centering
  \includegraphics[width=0.88\linewidth]{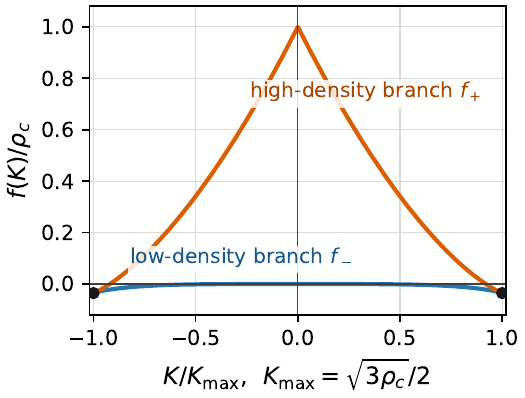}
  \caption{The two branches of the \(f(K)\) theory that reproduce the
  LQC ellipse, plotted in units \(\rhoc=1\).  The low-density branch
  \(f_-\) is analytic at \(K=0\) and starts at order \(K^4\).  The
  high-density branch \(f_+\) describes the neighborhood of the bounce
  at \(\rho=\rhoc\).  The branches meet at the maximum allowed
  \(|K|=\sqrt{3\rhoc}/2\).}
  \label{fig:fk_branches}
\end{figure}

Eliminating the auxiliary angle in favor of the mean curvature displays
the result directly in preferred-foliation variables and makes its branch
structure unavoidable.  Following the \(f(K)\) construction  in \cite{Afshordi:2010de}, one may write
\begin{align}
  S_f&=S_{\rm GR}+\int\dd^4x\sqrt{-g}\,f(K)
  \notag\\
     &=S_{\rm GR}
     -\int\dd^4x\sqrt{-g}\,[\chi K+U(\chi)].
  \label{eq:fk_aux}
\end{align}
The auxiliary representation is related to \(f(K)\) by
\begin{equation}
  K=-\frac{\dd U}{\dd\chi},
  \qquad
  f(K)=-\chi K-U(\chi),
  \qquad
  f_K=-\chi .
  \label{eq:legendre_fk}
\end{equation}
Here
\begin{equation}
  U[\chi(\theta)]
  =-\rhoc\sin^4\frac{\theta}{2},
  \qquad
  K=\frac{\sqrt{3\rhoc}}{2}\sin\theta .
  \label{eq:fk_theta}
\end{equation}
The map from \(K\) to density is two-branched.  On each monotonic leg,
terms linear in \(K\) shift the action by a boundary term.  Choosing these
integration constants separately on time-reversed legs gives a compact
even representation.  Define
\begin{equation}
  z=\frac{2|K|}{\sqrt{3\rhoc}},
  \qquad
  Q=\sqrt{1-z^2},
  \qquad 0\le z\le 1 .
\end{equation}
The low-density branch is
\begin{equation}
  f_-(K)=\rhoc\left[
    \frac{(1-Q)(3+Q)}{4}
    -\frac{z}{2}\arcsin z
  \right],
  \label{eq:fminus}
\end{equation}
and the high-density branch is
\begin{equation}
  f_+(K)=\rhoc\left[
    \frac{(1+Q)(3-Q)}{4}
    -\frac{z}{2}\left(\pi-\arcsin z\right)
  \right].
  \label{eq:fplus}
\end{equation}
The branches meet at \(z=1\), where
\[
f_-=f_+=\rhoc(3-\pi)/4.
\]
The even high-density representation is continuous but cusped at
\(K=0\): its one-sided derivatives select the contracting and expanding
legs.  Thus \(f_+\) is intrinsically branchwise rather than a globally
\(C^1\), single-valued function of \(K\).
Both branches are shown in Fig.~\ref{fig:fk_branches}.
On the low-density branch,
\begin{equation}
  f_-(K)=-\frac{K^4}{27\rhoc}
  +O\!\left(\frac{K^6}{\rhoc^2}\right),
  \label{eq:fk_low_expansion}
\end{equation}
so the Einstein-Hilbert term is recovered at low curvature.

The functions \eqref{eq:fminus}--\eqref{eq:fplus} obey
\begin{align}
  K f_\pm'(K)-f_\pm(K)&=-\rhoc x_\pm^2,
  \notag\\
  x_\pm(K)&=\frac{1}{2}
  \left[1\pm\sqrt{1-\frac{4K^2}{3\rhoc}}\right].
\end{align}
Here \(x_-\) and \(x_+\) label, respectively, the low- and high-density
roots.  Since their definition implies
\(K^2/3=\rhoc x_\pm(1-x_\pm)\), the Hamiltonian constraint of the
\(f(K)\) theory,
\begin{equation}
  \frac{K^2}{3}=\rho-f+Kf_K,
\end{equation}
first fixes \(\rho=\rhoc x_\pm\) and therefore gives
\begin{equation}
  \frac{K^2}{3}
  =\rhoc x_\pm(1-x_\pm)
  =\rho\left(1-\frac{\rho}{\rhoc}\right).
\end{equation}

\section{Discussion}

\subsection{Perturbations: cuscuton versus mimetic gravity}

  The original cuscuton is degenerate and carries
no local propagating fields \cite{Afshordi:2006ad}; formally it is the
\(c_s^2\to\infty\) limit, with an elliptic constraint on each preferred
slice rather than a wave equation.  A homogeneous CMC reduction may retain
one global mode and does not commute smoothly with construction of the
local phase space \cite{Gomes:2017tzd}, but at quadratic order the
pure-gravity effective cuscuton sector contains only the two tensor modes
\cite{Mylova:2023rqr}.  In explicit cuscuton-bounce models, the background
passes regularly through the bounce while the  cuscuton perturbations are algebraically solved for in Fourier space, leaving the comoving curvature perturbation of ordinary
matter as the physical scalar variable
\cite{Boruah:2018pvq,Boruah:2017tvg}; see also the
limiting-curvature analysis of Ref.~\cite{Quintin:2019limiting}.

Mimetic gravity lies at the opposite sound-speed limit.  In the minimal
theory its constraint supplies an irrotational pressureless scalar with
\(c_s^2=0\) and no ordinary spatial-gradient term
\cite{Arroja:2015wpa}.  Consequently there is no conventional weakly
coupled wave regime controlled by a finite sound horizon.  Analyses of
extended mimetic theories find gradient or ghost instabilities, while the
degenerate zero-gradient case instead signals strong coupling
\cite{Langlois:2018jdg,Takahashi:2017pje}.  Nonlinearly, the associated
geodesic potential flow generically develops caustics
\cite{Barvinsky:2013mea}.
Therefore, zero sound speed does not remove the mimetic scalar's
independent Cauchy data; the regular cuscuton instead supplies a
constraint with no additional local scalar initial data.

The \(F(\Box T)\) limiting-curvature operators modify this perturbation
system, but do not convert the mimetic scalar into a cuscuton.  Indeed, the
model reproduces the LQC background while its scalar perturbations differ
from the deformed-algebra LQC prescription \cite{deHaro:2018sqw}.
For this limiting-curvature reconstruction, the mimetic scalar
has  \(c_{s,{\rm mim}}^2=-2\rho/(3\rhoc)<0\), away from zero density
\cite{deHaro:2018sqw}; the zero-sound-speed statement applies to the
minimal mimetic theory. The
contrast
\begin{equation}
  c_{s,{\rm mim}}^2\,\leq0 {~~~~~\rm versus~}
  \qquad c_{s,{\rm cusc}}^2\to\infty,
  \label{eq:sound_speed_contrast}
\end{equation}
illustrates the distinction for the minimal and LQC-reconstructing
limiting-curvature mimetic models considered here. More generally,
the additional mimetic scalar, absent in the cuscuton, obstructs an
identification of the full theories through the homogeneous Legendre map.

Existing cuscuton-bounce calculations find a strongly blue
single-field scalar spectrum, while spectator entropy perturbations can
be scale invariant; tensor modes remain stable but typically blue
\cite{Kim:2020aem}.

\subsection{\texorpdfstring{Covariant completions and LQC
perturbations}{Covariant completions and LQC perturbations}}

Although the present construction reproduces the modified LQC background Friedmann equation, the corresponding perturbation theory is not uniquely determined by the background alone. This is a generic feature of effective cosmological theories: once the background evolution is specified, there remains freedom in how the perturbative degrees of freedom are introduced, truncated, and quantized. In effective LQC this is reflected by the dressed-metric \cite{Agullo:2012sh,Agullo:2012extension}, hybrid \cite{FernandezMendez:2012vi}, and deformed-algebra \cite{Cailleteau:2011kr} approaches, which share the same effective background dynamics but differ in their treatment of perturbations. The dressed metric approach describes the dynamics of  perturbations  on an effective quantum-corrected background geometry, whereas the hybrid approach is based on a canonical quantization of the perturbed cosmological system which is truncated at quadratic perturbative order in the action,  while in the deformed-algebra approach quantum corrections are incorporated in the constrained Hamiltonian system, requiring the absence of anomalies. Likewise, the lattice construction derives anomaly-free holonomy-corrected scalar constraints \cite{WilsonEwing:2012pu}, whereas the separate-universe approach follows long-wavelength perturbations across the bounce \cite{WilsonEwing:2015sfx}.  In the dressed-metric and hybrid approaches, the quantization is carried out by combining  LQG-inspired techniques for the background and Fock quantization for the perturbations. A detailed comparison of the dressed-metric and hybrid approaches is given in Ref.~\cite{Li:2022mln}.
~\\
The situation is different when a complete 4D covariant effective theory is available. In that case the classical perturbation equations are obtained uniquely by expanding the 4D covariant action around a chosen background, so that the dynamics of the perturbations are fixed by the underlying model. Remaining ambiguities are then associated only with the subsequent choice of variables, gauge and quantization of the perturbations rather than with their effective classical equations. This strategy  has recently been discussed, for example, in the context of generalised extended mimetic gravity models in \cite{Giesel:2025kdl}, where a fully covariant effective action provides the starting point for a consistent perturbative analysis beyond the homogeneous or spherically symmetric sectors.  
~\\
More generally, several inequivalent 4D covariant theories may reproduce the same homogeneous or even spherically symmetric effective dynamics. Agreement with additional symmetry reductions, such as cosmological perturbations or spherical symmetry perturbations, therefore provides increasingly stringent consistency checks but does not by itself establish uniqueness. The viability of a proposed covariant completion should be judged by the  extent to which it consistently reproduces independently derived effective LQG-inspired results across different symmetry sectors and their associated constraint structures.
For the angular action, we defer the explicit perturbation
calculation and its comparison with these LQC prescriptions. Stability
results for other cuscuton potentials cannot substitute for that test.
Agreement with effective anisotropic LQC is likewise not established by
the flat-FLRW reconstruction.

\subsection{Scope of the reconstruction}

The reconstruction is exact at the level of the homogeneous constraint.
The compact field \(\theta\) is the LQC polymerization angle, the constraint
fixes \(\rho=\rhoc\sin^2(\theta/2)\), and the negative cuscuton energy
supplies the \(-\rho^2/\rhoc\) term.  Because no equation of state enters,
the result holds for  minimally coupled matter sectors with a conserved
stress tensor and \(\rho+p>0\) on the regular branch.

The two reformulations clarify what kind of effective description this
is.  The mimetic Legendre transformation explains the angular coefficient
and potential, but only after the clock and cuscuton gradients are aligned
homogeneously.  Higher-derivative mimetic actions that reproduce open and
closed LQC backgrounds \cite{Langlois:2017hlu} therefore do not follow
automatically from the flat angular action.  The \(f(K)\) form instead
exposes the preferred CMC foliation.  It shares this language with the
low-energy cuscuton/Ho\v rava--Lifshitz correspondence
\cite{Horava:2009uw,Afshordi:2009tt}, but is a branched function tailored
to the LQC ellipse, not the quadratic cuscuton.

\section{Conclusions}
We have shown that the flat-FLRW LQC Friedmann equation is the homogeneous
constraint of an angular cuscuton theory.  With
\(\theta=2\lambda b\), the square-root action
\eqref{eq:theta_cuscuton} gives
\(H=(1/2)\sqrt{\rhoc/3}\sin\theta\) and
\(\rho=\rhoc\sin^2(\theta/2)\), including the bounce at
\(\theta=\pi\). The construction is closed-form, local, generally
covariant, valid for  minimally coupled matter with
\(\rho+p>0\) on the regular branch, and introduces no
local propagating fields.

From the first-order perspective, the constant-tension cuscuton
coordinate is the Legendre-dual variable
\(\Theta=P=\sin\theta-\theta\), with canonical normalization
\(\chi=-\sqrt{\rhoc/3}\,\Theta\).  It is related nonlinearly to the
polymerization-angle coordinate \(\theta=2\lambda b\), and the two
parametrizations yield the same homogeneous equations of motion.

The mimetic Legendre map explains the special angular coefficients, while
the explicit branched \(f(K)\) form places the same dynamics in the
preferred-foliation language associated with cuscuton and Ho\v
rava--Lifshitz gravity.  These links are exact for the homogeneous
constraint but do not identify the perturbations of the respective
theories.

Whether the equivalence established at the level of the modified Friedmann  equation and its Hamiltonian formulation extends beyond homogeneous cosmology remains an interesting open question. The distinct perturbative properties of the  mimetic-gravity and cuscuton models suggest that a straightforward extension of the  Legendre and canonical transformations presented here is unlikely. Nevertheless, a more general correspondence between other  symmetry-reduced sectors or the full 4D covariant theories, possibly involving a singular transformation, cannot be excluded and  merits future investigation. From a phenomenological standpoint, however, the absence of evidence for additional gravitational degrees of freedom, together with the observed stability of the local vacuum, provides empirical motivation for our cuscuton representation of LQC cosmology.

\begin{acknowledgments}
NA thanks Ghazal Geshnizjani and Edward Wilson-Ewing for useful
discussions.  We thank the Banff International Research Station (BIRS),
the Institute for Advanced Study in Mathematics (IASM), and the organizers
and participants of the workshop ``Recent Development of Quantum Gravity
and Applications to Cosmology and Black Hole Physics'' (26w5588), held in
Hangzhou, China, for their hospitality and stimulating discussions.  The
bulk of this work was carried out during the workshop.  NA is supported by
the Natural Sciences and Engineering Research Council of Canada (NSERC)
and the Perimeter Institute for Theoretical Physics.  Research at Perimeter
Institute is supported in part by the Government of Canada through the
Department of Innovation, Science and Economic Development and by the
Province of Ontario through the Ministry of Colleges and Universities.
KG thanks Perimeter Institute for Theoretical Physics for hospitality and
support.  KG was supported by a grant from the Simons Foundation (1034867,
Dittrich). The authors acknowledge use of  OpenAI Codex and Anthropic Claude, to assist with transcribing handwritten derivations, literature searches, algebraic consistency checks, and preparation of plotting code. The authors directed the use of these tools, reviewed and verified its outputs, and take full responsibility for the content of this paper.
\end{acknowledgments}

\end{document}